\documentclass[prb,preprint,superscriptaddress]{revtex4}

\usepackage{graphicx}

\begin{document}

\title{Superconductivity and quantum criticality in heavy fermions CeIrSi$_3$ and CeRhSi$_3$}


\author{J. F. Landaeta}
\altaffiliation{Also at Departamento de F\'{\i}sica, Facultad de Ciencias, Universidad Central de Venezuela, Apartado 47586, Caracas 1041-A, Venezuela}
\author{D. Subero}
\author{D. Catal\'{a}}
\altaffiliation{Also at Departamento de F\'{\i}sica, Facultad de Ciencias, Universidad Central de Venezuela, Apartado 47586, Caracas 1041-A, Venezuela}
\author{S. V. Taylor}
\altaffiliation{Now at Cavendish Laboratory, Cambridge University, JJ Thomson Avenue, Cambridge CB3 OHE, U.K.}
\affiliation{Centro de F\'{\i}sica, Instituto Venezolano de
		Investigaciones Cient\'{\i}ficas, Apartado 20632, Caracas
		1020-A, Venezuela}

\author{N. Kimura}
\affiliation{Department of Physics, Graduate School of Science, Tohoku University, Sendai 980-8578, Japan}

\author{R. Settai}
\affiliation{Department of Physics, Niigata University, Niigata 950-2181, Japan}

\author{Y. \={O}nuki}
\affiliation{Faculty of Science, University of the Ryukyus, Nishihara, Okinawa 903-0213, Japan}

\author{M. Sigrist}
\affiliation{Institute for Theoretical Physics, ETH Z\"{u}rich, CH-8093, Z\"{u}rich, Switzerland}

\author{I. Bonalde}
\affiliation{Centro de F\'{\i}sica, Instituto Venezolano de
	Investigaciones Cient\'{\i}ficas, Apartado 20632, Caracas
	1020-A, Venezuela}

\begin{abstract}
\textbf{Superconductivity and magnetism are mutually exclusive in most alloys and elements, so it is striking that superconductivity emerges around a magnetic quantum critical point (QCP) in many strongly correlated electron systems (SCES). In the latter case superconductivity is believed to be unconventional and directly influenced by the QCP. However, experimentally unconventional superconductivity has neither been established nor directly been linked to any mechanism of the QCP. Here we report measurements in the heavy-fermion superconductors CeIrSi$_3$ and CeRhSi$_3$. The measurements were performed with a newly developed system, first of its kind, that allows high-resolution studies of the superconducting gap structure under pressure. Superconductivity in CeIrSi$_3$ shows a change from an excitation spectrum with a line-nodal gap to one which is entirely gapful when pressure is close but not yet at the QCP. In contrast, CeRhSi$_3$ does not possess an obvious pressure-tuned QCP and the superconducting phase remains for all accessible pressures with a nodal gap. Combining both results suggests that unconventional behaviours may be connected with the coexisting antiferromagnetic order. This study provides a new viewpoint on the interplay of superconductivity and magnetism in SCES.}
\end{abstract}

\maketitle

Magnetic order, antiferromagnetism or ferromagnetism, is a rather common feature in a wide range of SCES, such as high-T$_c$ cuprates\cite{Moriya2000}, organics\cite{Kuroki2006}, iron pnictides/chalcogenides\cite{Abrahams2011}, and heavy fermion materials\cite{Pfleiderer2009,White2015}. Intriguingly, the weakening of magnetic order by some tuning parameters, such as pressure, chemical substitution or magnetic fields, promotes the emergence of superconductivity mostly linked to a QCP where the transition temperature for the magnetic order (antiferromagnetic order in our study) vanishes. In a few cases, such as the hole-doped cuprates, the antiferromagnetic (AFM) phase disappears before superconductivity rises\cite{Moriya2000}. The more common situation, however, is a ''superconducting dome'' around the QCP, where $T_c$ takes a maximal value near the QCP. This implies that before the N\'eel temperature $T_N$ vanishes superconductivity and AFM coexist in roughly half of the dome region.

In most heavy-fermion compounds the presence of a QCP agrees well with the famous Doniach phase diagram for Kondo lattice systems, where Kondo singlet screening competes with the ordering of localized magnetic moments. Beyond this, the physics associated with the coexistence and the possible competition of antiferromagnetism and superconductivity is highly intriguing. A long-standing question is the microscopic understanding of the mechanism that supports unconventional superconductivity around the QCP. In many heavy-fermion systems, including CeIrSi$_3$ and CeRhSi$_3$, it is inferred that Kondo-singlet correlation enhances quantum spin fluctuations near the QCP.

Experimental work to explore the details of the phase diagrams has been scarce, mainly because of the technical challenges involved with the doping of some SCES or the implementation of any of the common magnetic or thermodynamic probes at high pressures and very low temperatures. Regarding the superconducting phase, the order parameter symmetry, determining the energy gap structure as signature for conventional or unconventional pairing, has been addressed across the dome for only few SCES. Pressure tuning in most heavy fermions is an advantage as compared to the complexity of chemical substitution in high-$T_c$ cuprates or iron pnictides/chalcogenides. Doping can introduce disorder, and the intrinsic effects originating from the quantum phase transition can be masked as a result of the disorder sensitiveness of quantum fluctuations. Our present study aim at gain more insight into the behavior of superconductivity around a QCP by studying the properties of energy gap structure for the non-centrosymmetric compounds CeIrSi$_3$ and CeRhSi$_3$.

CeIrSi$_3$ orders antiferromagnetically below the Neel temperature $T_N=5.0$ K at ambient pressure\cite{Sugitani2006}. As pressure is applied $T_N$ decreases monotonically toward zero. Superconductivity has its onset around 1.3 GPa and persists up to 3.5 GPa with a maximum critical temperature around 1.6 K at $p_c^*\approx2.58$ GPa \cite{Tateiwa2007}. So far, the antiferromagnetic order has not been detected directly inside the superconducting phase above the critical pressure $p_c=2.25$ GPa, at which $T_N=T_c$. An extremely high upper critical field of 45 T for $H \parallel [001]$ at 2.6 GPa \cite{Settai2008} and a possible nodal energy gap at $p=2.8$ GPa higher than $p_c^*$ \cite{Mukuda2008a} were reported previously. A slightly different situation is encountered in CeRhSi$_3$ which becomes an antiferromagnet below $T_N=1.6$ K at ambient pressure\cite{Kimura2005}. Upon increasing pressure $T_N$ passes through a maximum of 1.9 K and then decreases smoothly and approaches asymptotically the superconducting dome at its maximum at $p_c^*\approx2.8$ GPa. Thus, $ T_N $ of the AFM phase of CeRhSi$_3$ seems not head towards zero as expected for a QCP, in contrast to the observation in CeIrSi$_3$. Very high upper critical fields were also found in CeRhSi$_3$, exceeding 25 T for $H\parallel[001]$ at 2.85 GPa\cite{Sugawara2010}. It is also important to note here that CeIrSi$_3$ and CeRhSi$_3$ exhibit itinerant $f$-electrons, i.e. heavy-fermion behavior, at ambient pressure\cite{Aso2007,Ohkochi2009,Aso2012} and keep similar Fermi surfaces\cite{Terashima2008,Onuki2012} over all the pressure range\cite{Terashima2007,Yamaoka2011}.

The lack of inversion symmetry in these compounds leads to the appearance of an antisymmetric spin-orbit coupling (ASOC) and to the mixing of spin-singlet and spin-triplet superconducting states. This does, however, not preclude that ASOC is responsible for unconventional superconductivity, which is supposed to originate from non-phonon mediated pairing. This is obvious in the case of the isostructural CePt$_3$Si and LaPt$_3$Si; where the former is a heavy-fermion system and shows unconventional superconductivity, while the latter is rather an ordinary  superconductor with electron-phonon mediated Cooper pairing\cite{Ribeiro2009}.

Here we report a systematic study on the temperature dependence of the magnetic penetration depth $\lambda(T)$  of CeIrSi$_3$ and CeRhSi$_3$ under hydrostatic pressure. The penetration depth is widely considered one of the most sensitive probes to detect the nodal structure of the superconducting energy gap, which is related to the order parameter symmetry. The high-pressure low-temperature study required the development of a completely new experimental setup (see Methods).
In the normal state above $T_c$, which we define as the onset of the diamagnetic response, our experimental setup probes the skin depth; i.e., $f(T)\propto\delta(T)=\sqrt{2\rho(T)/\mu(T)\omega}$. Here $\rho(T)$ is the resistivity, $\mu(T)$ is the permeability, and $\omega=2\pi f$ is related to the oscillator frequency $f$. In the case of the magnetic superconductors CeIrSi$_3$ and CeRhSi$_3$, below $T_c$ the effective penetration depth is given by $\lambda(T)=\sqrt{\mu(T)}\lambda_L(T)$, where $\lambda_L(T)=\sqrt{n_0/n_s(T)} c/\omega_p$. Here $\omega_p$ is the plasma frequency, $c$ is the speed of light, $n_0$ is the carrier density and $n_s$ is the superconducting electron density. In the low-temperature limit $\mu(T)$ of antiferromagnets approaches 1 exponentially\cite{Lidiard1954,Fawcett1988}, such that then we directly probe the Cooper-pair density as $f(T)\propto\lambda(T)=\lambda_L(T)$.

\begin{figure}\begin{center}
		\scalebox{1}{\includegraphics{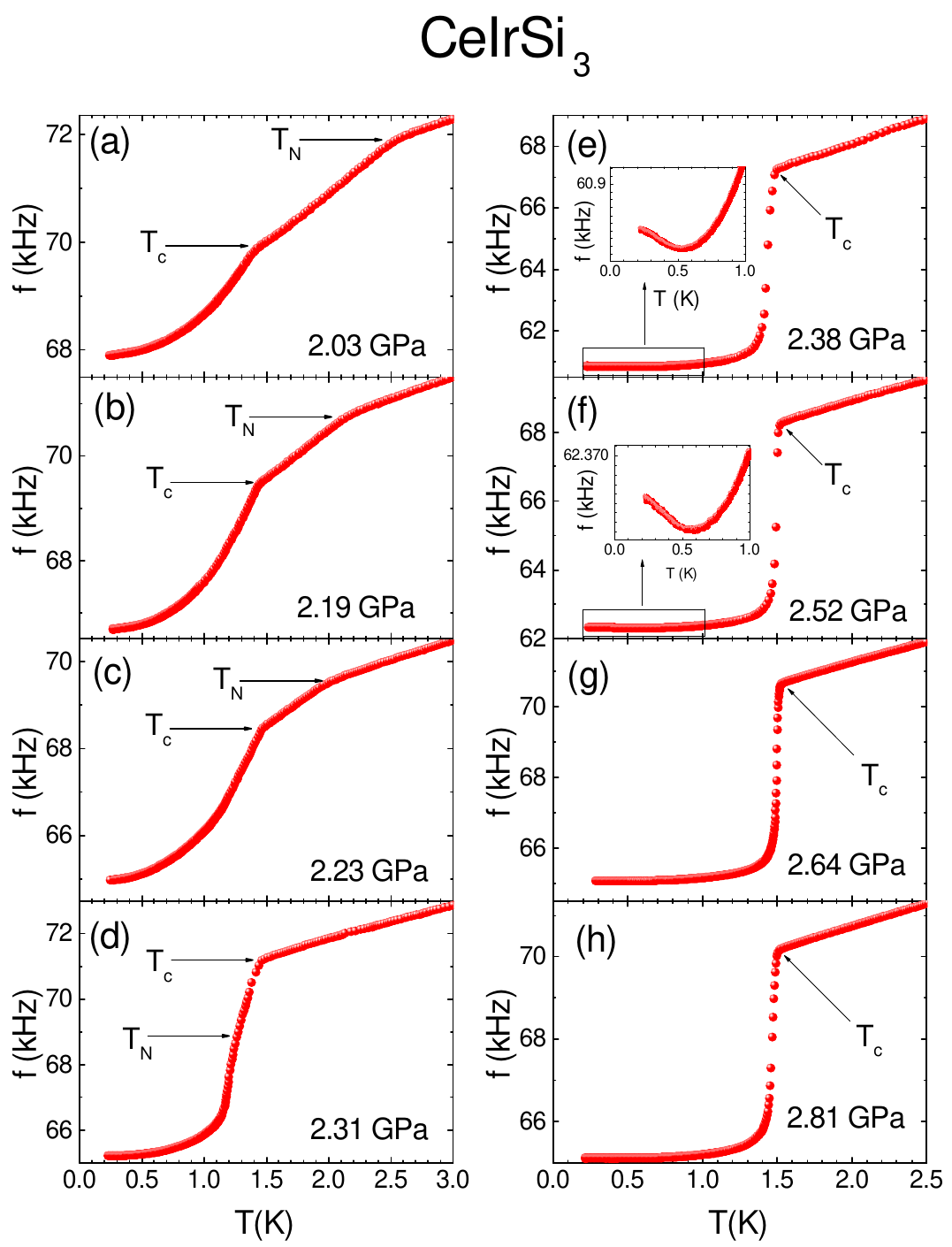}}
		\caption{\label{fig:Figure_1}{\textbf{Magnetic penetration depth and skin depth of CeIrSi$_3$ under pressure.} Temperature dependence of the oscillation frequency, proportional to the penetration depth at $T<T_c$ and the skin depth at $T>T_c$, for different pressures. $T_c$ marks the onset of the superconducting transitions and $T_N$ is associated with AFM transitions, in accordance with previous measurements. The insets of \textbf{e} and \textbf{f} show upturns around 0.4 K related to pair-breaking paramagnetic precursors of an AFM transition.}}\end{center}
\end{figure}

High-resolution measurements were carried out for both materials on single crystals down to 200 mK and up to 2.81 GPa. Figure~\ref{fig:Figure_1}a-h show the oscillation frequency variation $f(T)$ for CeIrSi$_3$ at different pressures. Apart from the superconducting transitions shown by the penetration depth and indicated as $T_c$ in Fig.~\ref{fig:Figure_1}a-h, the skin depth displays a marked change of slope at points labelled as $T_N$ in Fig.~\ref{fig:Figure_1}a-c. We attribute these anomalies to AFM transitions, as the change occurs in each case near the reported $T_N$ \cite{Sugitani2006}. We remark that these anomalies in the skin depth can reflect the temperature response of resistivity and permeability.

\begin{figure}\begin{center}
		\scalebox{1}{\includegraphics{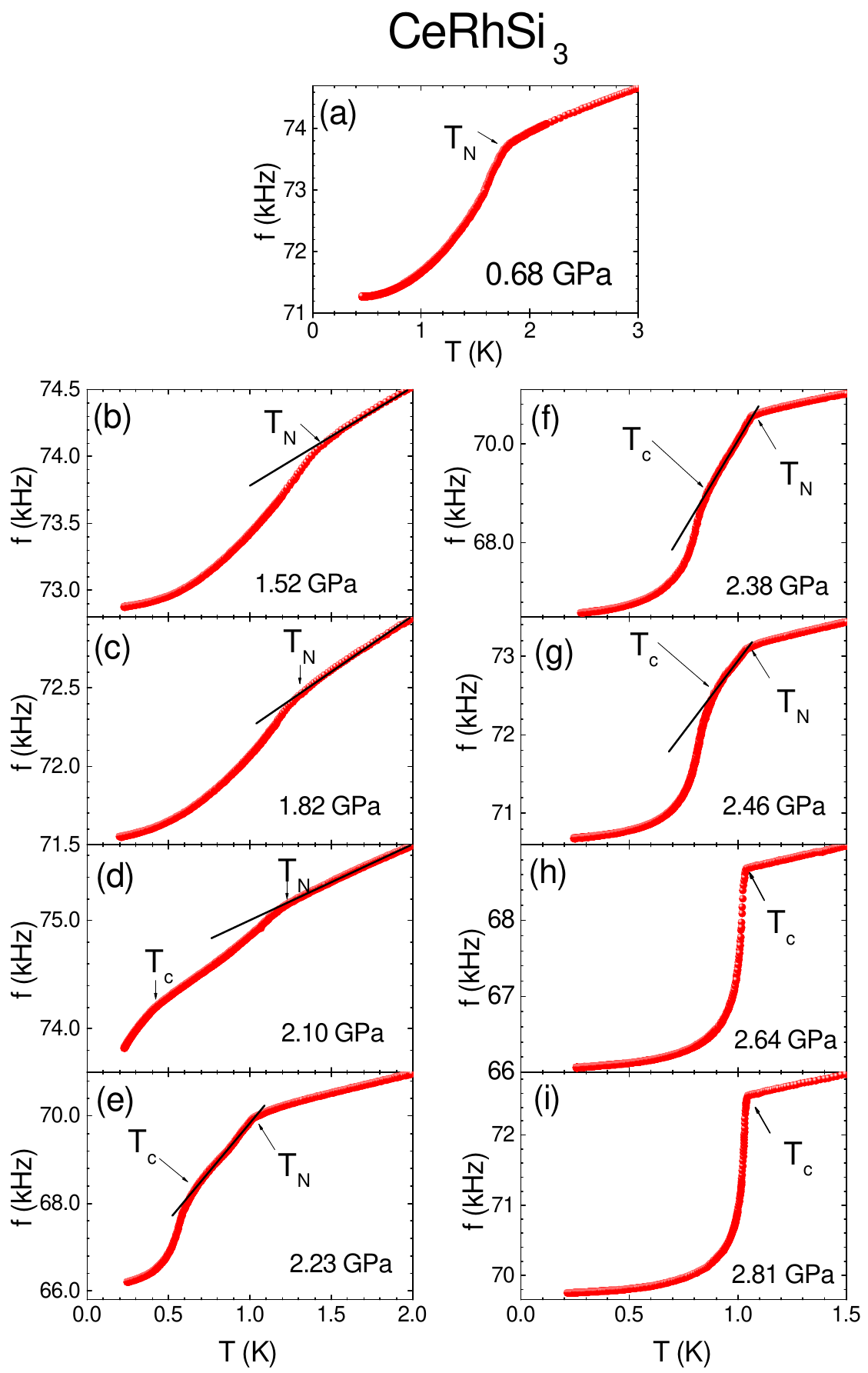}}
		\caption{\label{fig:Figure_2}{\textbf{Magnetic penetration depth and skin depth of CeRhSi$_3$ under pressure.} Temperature dependence of the oscillation frequency for different pressures. $T_c$ marks the onset of the superconducting transitions and $T_N$ is associated with AFM transitions, following previous measurements. No sign of antiferromagnetism is observed below $T_c$ in \textbf{h} and \textbf{i}.}}\end{center}
\end{figure}

Our measurements in CeIrSi$_3$ also show for the first time signs of the AFM phase when $T_N<T_c$. We argue that the bump at the transition for 2.31 GPa (Fig.~\ref{fig:Figure_1}d) and the upturns observed at very low temperatures for 2.38 and 2.52 GPa (Fig.~\ref{fig:Figure_1}e,f) are likely signatures of the low-temperature AFM order or fluctuations near $p_c^*$ \cite{Jacobs1995,Chia2001}.
This detection of the AFM phase for $T_N<T_c$ ($p_c<p<p_c^*$) is in line with direct evidence found early in other heavy fermion systems\cite{Park2006,Egetenmeyer2012,Shen2016}. In CeRhIn$_5$, for example, the AFM phase was detected by specific-heat measurements after the quenching of superconductivity with high magnetic fields\cite{Park2006}. Figure~\ref{fig:Figure_2}a-i display $f(T)$ for CeRhSi$_3$ at different pressures. The skin depth exhibits slope changes at points indicated by $T_N$ in Fig.~\ref{fig:Figure_2}a-g, which coincide with previously reported AFM transitions in CeRhSi$_3$ \cite{Kimura2005}. Contrary to the observation in CeIrSi$_3$, no trace of an AFM phase is detected below $T_c$ in CeRhSi$_3$.

\begin{figure}\begin{center}
		\scalebox{1}{\includegraphics{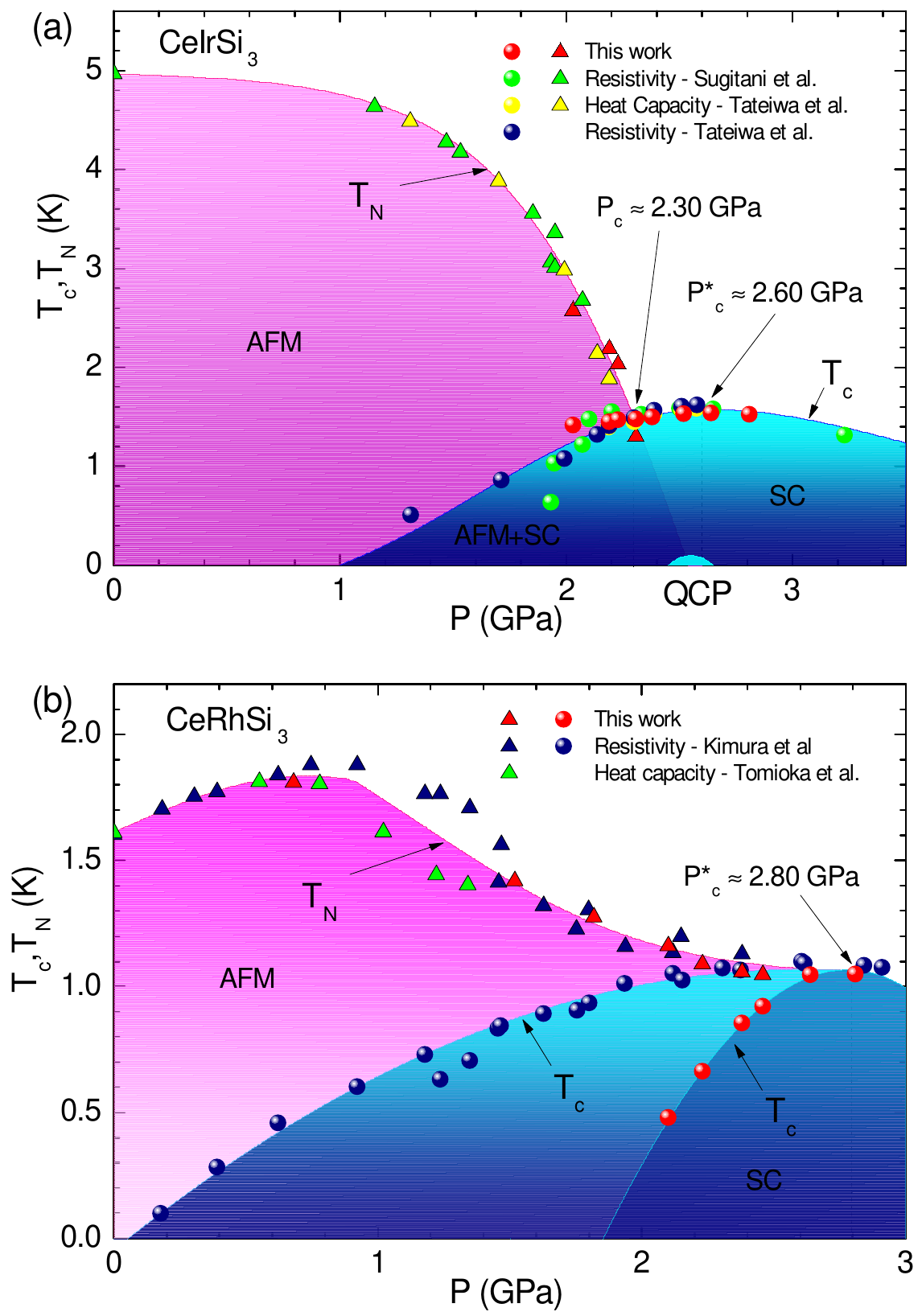}}		\caption{\label{fig:Figure_3}{\textbf{Temperature-pressure phase diagrams of CeIrSi$_3$ and CeRhSi$_3$.} Phase diagrams of CeIrSi$_3$ and CeRhSi$_3$ composed of our and some previously reported data.  N\'eel (triangles) and superconducting (spheres) critical temperatures as obtained by resistivity, specific heat, and the present work. \textbf{a} Within experimental error the QCP in CeIrSi$_3$ coincide with the pressure $p_c^*$ at which $T_c$ is the highest and superconductivity is optimal. \textbf{b} Notoriously, there seems to be no AFM QCP in CeRhSi$_3$, as the AFM phase appears to vanish asymptotically at the superconducting dome maximum. Our $T_c$ data agree with recent values from heat-capacity measurements\cite{Umehara2016}.}}\end{center}
\end{figure}

We used our own and reported values of $T_c$ and $T_N$\cite{Sugitani2006,Tateiwa2007,Kimura2007,Tomioka2007} to construct the $T-p$ phase diagrams of CeIrSi$_3$ and CeRhSi$_3$ displayed in Fig.~\ref{fig:Figure_3}a,b. The general agreement of the CeIrSi$_3$ data is remarkable, whereas no concurrence in the $T_c$ values is found in CeRhSi$_3$. The large data scattering for $p<2.0$ GPa in CeIrSi$_3$ is due to difficulties in defining the critical temperatures from the very broad transitions occurring at those pressures . On the other hand, we think that the inconsistency in CeRhSi$_3$ may be caused by the possible existence of surface superconductivity sensed by resistivity measurements at low pressures, as has been suggested for CeIrSi$_3$\cite{Iida2016}. We cannot discard sample issues. Very recent specific-heat measurements found no sign of superconductivity below 1.8 GPa\cite{Umehara2016}. It is worth noting that the low-pressure uncertainties in $T_c$ do not affect the conclusions of the present study. For CeIrSi$_3$ we identify in Fig.~\ref{fig:Figure_3}a the critical pressures $p_c\approx2.30$ GPa and $p_c^*\approx2.60$ GPa. At the latter pressure antiferromagnetism vanishes (QCP), $T_c$ is maximum, and several other properties indicate that superconductivity is optimal. Both critical pressures are very close to the values reported early\cite{Sugitani2006,Tateiwa2007,Settai2011}.

The AFM phase behaves quite remarkably in CeRhSi$_3$ (Fig.~\ref{fig:Figure_3}b). The phase boundary lines of the AFM and superconducting phases do not cross but seem to merge at a pressure close to $p_c^*\approx2.80$ GPa. Based on our observation there is no signature for a pressure-tuned QCP in CeRhSi$_3$, at least in the pressure range studied. We could not detect any distinct sign of an AFM transition above 2.6 GPa. This contrasts with muon-spin rotation spectroscopy, which sensed an internal magnetic field in single crystals of CeRhSi$_3$ down to its disappearance at very low temperature \cite{Egetenmeyer2012}. This internal field was assumed to be associated with antiferromagnetism and the result was interpreted as evidence of a QCP in CeRhSi$_3$. In a different study the AFM phase shows up again under the application of a magnetic field of 4 T at 2.61 GPa \cite{Iida2013}, suggesting that indeed the AFM order is hidden or suppressed by superconductivity at zero field.

\begin{figure}\begin{center}
		\scalebox{0.95}{\includegraphics{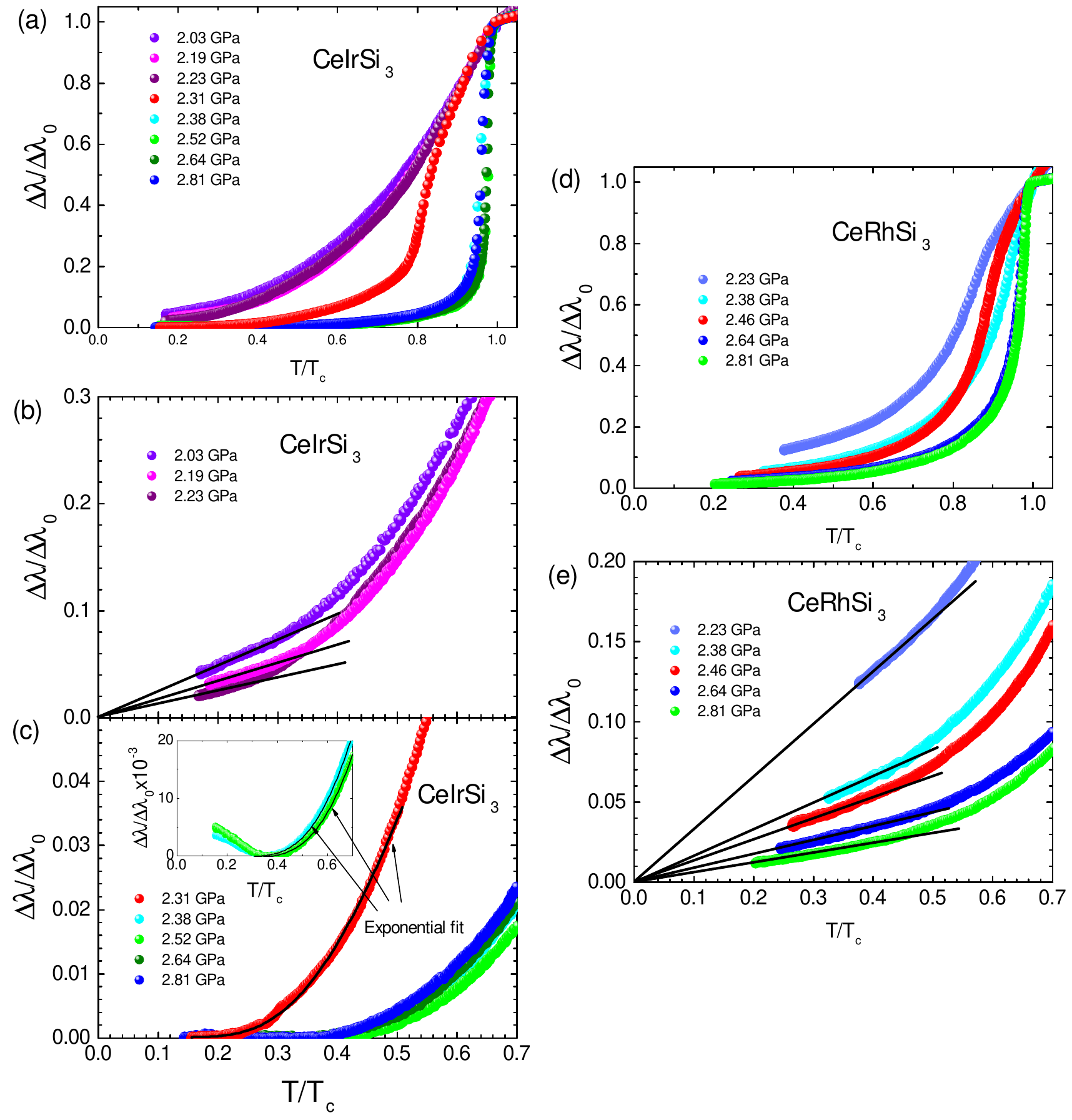}}
		\caption{\label{fig:Figure_4}{\textbf{Magnetic penetration depth of CeIrSi$_3$ and CeRhSi$_3$ under pressure}. \textbf{a,d} Magnetic penetration depth normalized to the total penetration-depth shift $\Delta \lambda_0$ clearly shows a continuous reduction of superconducting transition width with pressure in both CeIrSi$_3$ and CeRhSi$_3$. \textbf{b} Low-temperature penetration depth of CeIrSi$_3$ shows a linear behaviour for pressures lower than $p_c=2.30$ GPa, at which the AFM order becomes significantly weak. \textbf{c} The penetration depth of CeIrSi$_3$ changes to an exponential response for $p > p_c$. \textbf{e} Low-temperature linear response of the penetration depth of CeRhSi$_3$ at all pressures studied. As opposed to the observation in CeIrSi$_3$, no crossover to exponential response is detected.}}\end{center}
\end{figure}

Now we discuss in detail the superconducting phase of both compounds, starting with CeIrSi$_3$. The data in Fig.~\ref{fig:Figure_1}a-h are limited to the superconducting region and converted to penetration depth, they are displayed in Fig.~\ref{fig:Figure_4}a. Low-temperature regions are expanded in Fig.~\ref{fig:Figure_4}b,c for pressures around the center of the superconducting dome. A conspicuous change of the qualitative behaviour of the penetration depth can be observed, both at the transition and at low temperatures, near the pressure $p_c=2.30$ GPa.  This pressure is distinctively lower than the one defining the AFM QCP ($\sim$2.6 GPa) and, remarkably, no qualitative change in the temperature dependence of the penetration depth is seen at the  QCP. The most striking result concerns  the low-temperature behavior. Below $p_c$ the penetration depth displays a linear temperature dependence (Fig.~\ref{fig:Figure_4}b), which indicates the presence of gapless quasiparticle excitations due to line nodes in the superconducting energy gap. The slope fades out as $p\rightarrow p_c$ and the penetration depth changes to an exponential temperature dependence (Fig.~\ref{fig:Figure_4}c), a characteristic behaviour signalling the existence of a nodeless superconducting energy gap. At all pressures above $p_c$ this exponential dependence is seen (the AFM-caused upturns at 2.38 and 2.52 GPa were subtracted), even at the QCP where no further variation occurs. It is important to note here that our results contradict the interpretation of $^{29}$Si NMR data in terms of line nodes for pressures around 2.7-2.8 GPa \cite{Mukuda2008a}.

Near the QCP of CeIrSi$_3$ we found that $T_c$ is maximal (see phase diagram in Fig.~\ref{fig:Figure_3}a) and the superconducting transition is the narrowest (Fig.~\ref{fig:Figure_4}a), which implies that superconductivity is optimal at this point. The ratio $\Delta_0/k_B T_c$ is 3.14 and 3.55 for pressures 2.38 GPa and 2.52 GPa, respectively, then decreases to 2.35 for pressures above $p^*_c$. All the ratio values in the range of 2.38-2.81 GPa are higher than 1.76 expected from the BCS weak-coupling limit. This rather high values could be interpreted as strong-coupling corrections. The zero-temperature energy gap $\Delta_0$ was determined from the fit to the BCS exponential approximation at low temperatures $\Delta\lambda(T)\propto\sqrt{\pi\Delta_0/2k_BT} \exp(-\Delta_0/k_BT).$ Here $k_B$ is the Boltzmann constant. All findings are in agreement with early resistivity and heat-capacity results\cite{Tateiwa2007}.

Considering CeRhSi$_3$, and following the same procedure as for CeIrSi$_3$, the magnetic penetration depth is presented in Fig.~\ref{fig:Figure_4}d, with the low-temperature region zoomed in Fig.~\ref{fig:Figure_4}e. The superconducting transition sharpens as pressure increases up to $p_c^\star$, where $T_c$ is maximal, much similar as found for CeIrSi$_3$. However, the penetration depth displays a linear temperature dependence, which implies the presence of line nodes in the superconducting energy gap of this compound up to the highest applied pressure. Notably, the slope gradually gets smaller as pressure increases, even slightly above $p_c^\star$.

Based on our findings we may state that in CeIrSi$_3$ and CeRhSi$_3$ nodal superconductivity only appears in regions where $T_N\geq T_c$; i.e., where the magnetic dominates over the superconducting energy scale. The transition from a nodal to a nodeless superconducting energy gap at $p_c$ in CeIrSi$_3$ and the existence of line nodes in the absence of a QCP in CeRhSi$_3$ provide first experimental evidence that the nodal superconductivity in these compounds seems to be connected with the coexistence to a magnetic order. Another issue is that the superconductivity becomes strongest at $p_c^\star$ in both compounds, pressure at which the magnetic order vanishes in CeIrSi$_3$ but still appears to exist in CeRhSi$_3$. The values of the upper critical fields, the zero-temperature energy gaps, the heat-capacity jumps, among others, at $p_c^\star$ suggest that superconductivity in these two compounds is non-standard and not driven by electron-phonon interactions.

The implications of our investigation may be extended to other SCES in which superconductivity and magnetism coincide, specifically other heavy-fermion compounds. In particular, among the latter ones are other non-centrosymmetric materials of the family Ce$TX_3$ ($T$:transition metal; X:Si,Ge), CePd$_2$Si$_2$, some members of the series Ce$_n$M$_n$In$_{3n+2m}$ (CeIn$_3$, Ce$_2$RhIn$_8$, and CeMIn$_5$), and the $5f$-electron UCoGe (see the review by Pfleiderer for the heavy-fermion family\cite{Pfleiderer2009}). Another relevant compound is the noncentrosymmetric CePt$_3$Si\cite{kimura2012}, whose superconductivity coexists with antiferromagnetism from ambient pressure to about 0.6 GPa where the AFM phase vanishes at $T=0$ and also has line nodes confirmed by penetration depth measurements. Interestingly,  the tuning-parameter dependence of the superconducting energy gap for iron pnictides and CrAs do not show any anomalous variation in connection with the critical point where the AFM order seizes to be observable or at the QCP\cite{Hardy2010,Gordon2010,Hashimoto2012,Khasanov2015,Cho2016}. The difference with respect to CeIrSi$_3$ may be associated with the nature of the heavy quasiparticles in heavy-fermion systems whose magnetic properties are dominated by the nearly localized $4f$-moments. In contrast, in CrAs and iron-pnictides magnetic order evolves from much more itinerant $d$-electrons.

\textit{Interpretation of the superconducting energy gap structure}. The connection of the superconducting gap structure (line nodes) as well as the coexistence with dominant AFM order in both compounds is intriguing. A possible scenario to understand this property could be based on the fact that the underlying symmetries of the electronic system is different with and without AFM order. The lower symmetry of the AFM state may lead to a
reduced phase space for Cooper pairing favouring a nodal pairing state, while in the absence of AFM order Cooper pairs can take advantage of a enlarged set of possible pairing states enabling the condensate to avoid nodes. This viewpoint does not require a qualitative change of the pairing interaction which may originate from the same magnetic fluctuation associated with QCP. This reminds of the discussion of the superconducting double transition observed in UPt$_3$ which may be connected to the slight lifting of the degeneracy in a multi-component order parameter space due to AFM order, leading to "high"-temperature phase with more nodes than the "low"-temperature phase\cite{Joynt-Taillefer2002}. This type of scenario leads likely a superconducting phase which breaks time reversal symmetry
for the nodeless phase, a feature which could be probed by zero-field $ \mu$SR. Another setting to explain our findings can be based the lack of inversion symmetry.  AFM order combined with parity-mixture due to strong ASOC could cause the observed nodal gaps in these non-centrosymmetric superconductors, as suggested on theoretical grounds\cite{Fujimoto2006,Yanase2007}. These studies, however, ignore the
complex nature of the heavy-fermion physics.

\textit{Conclusions} We carried out a study on the temperature dependence of the magnetic penetration depth $\lambda(T)$ of CeIrSi$_3$ and CeRhSi$_3$ under hydrostatic pressure. In CeIrSi$_3$ we found an evolution from line-node to isotropic gap superconductivity at a pressure distinctively lower than the one at the QCP. In CeRhSi$_3$ no QCP was observed within our measurements and line nodes exist for all applied pressures. These results provide first evidence that the excitation gap of the superconducting phase for both compounds, CeIrSi$_3$ and CeRhSi$_3$, is closely connected with the antiferromagnetic order while the spin fluctuations favour to an unconventional pairing channel. More generally, our finding show that the interplay of superconductivity and magnetism in heavy-fermion materials and other SCES may offer more intriguing features which could be explored by other means.

\vspace{30pt}

\noindent {\large \textbf{Methods}}

\subsection{Sample preparation}

Single crystals of CeIrSi$_3$ were grown by the Czochralski method in a tetra-arc furnace with a pulling speed of 15 mm/h \cite{Sugitani2006}. The grown ingots were then wrapped in Ta-foil, sealed in a quartz tube under a vacuum of 10$^{-6}$ Torr and annealed at 950 $^\circ$C for five days. The high quality of the samples is evidenced by typical residual resistivity ratios of 120. The measured samples were cut to dimensions of $0.45 \times 0.65 \times 0.2$  mm$^3$ and polished with alumina of grain size about 0.3 $\mu$m.

Single crystals of CeRhSi$_3$ were grown by the Czochralski pulling method in a tetra-arc furnace under Ar gas atmosphere. Starting materials were 4N-Ce, 3N-Rh, and 5N-Si. The resulting ingots were annealed at 900 $^\circ$C under a vacuum of $2\times 10^{-6}$ Torr for a week\cite{Kimura2005}. The typical residual resistivity ratio was about 180. The measured samples were cut to dimensions of $0.5 \times 0.5 \times 0.24$  mm$^3$.

\subsection{Experimental setup}

Measurements were performed using a 13 MHz tunnel diode oscillator\cite{Bonalde2005} coupled to a self-clamped hybrid-piston-cylinder cell made of nonmagnetic CuBe, WC, and NiCrAl alloys (C\&T Factory Co., Ltd.). A 2.0-mm-diameter Cu-wire coil with a 7.2-mm winding height was placed into the 4.0-mm-diameter inner space of the cell filled with a pressure-transmitting medium. Samples were attached, with fast-bonding glue, to a tiny plastic rod fitted inside the coil.  The coil and the plastic rod were thermally anchored to the pressure cell, which was mounted to the mixing chamber of a dilution refrigerator. The samples were aligned with the $ab$-planes perpendicular to the probing magnetic field of less than 5 mOe, so we sensed the in-plane penetration depth. The variation of the measured frequency from its value at the lowest temperature, $\Delta f(T)$, is directly proportional to the penetration-depth ($\Delta\lambda(T)$) and skin-depth ($\Delta\delta(T)$) shifts below and above $T_c$, respectively\cite{Bonalde2005}. A thermometer was located close to the attaching point of the plastic rod at the pressure cell. No appreciable background signal was detected in the temperature range scanned. The experimental system was tested by measuring high-purity aluminum, tin, zinc,and cadmium samples; all runs yielded the expected BCS temperature dependence of $\lambda$ and critical temperature at different pressures.

A fiber-optic setup that measured the ruby R1 fluorescence line shifts was employed for estimating the pressure inside the piston-cylinder cell at 4 K. Ruby crystals were fixed, also with fast-bonding glue, to the end of the optical fiber and placed very near the sample inside the coil. Glycerol was chosen as the pressure-transmitting medium because of its better thermal conductivity properties at temperatures below 1 K. We determined pressure with the expression\cite{Yamaoka2012}
\begin{equation}
P(\mbox{GPa})= A_0\ln \left(\frac{\lambda}{\lambda_0}\right) \label{plt} \,,
\end{equation}
\noindent where $A_0=1762$ GPa and the wavelength $\lambda_0=693.33$ nm. Our pressure error was about 0.04 GPa.

\vspace{30pt}

\noindent{\large \bf References}


\subsection{Acknowledgements} We appreciate assistance from F. Honda at the early stage of the pressure project. We value conversations with S. Saxena. We acknowledge support from the Venezuelan Institute for Scientific Research (IVIC) grant No. 441. The work by N.K. was supported by JSPS KAKENHI grant No. JP26400345.

\subsection{Author Contributions} J.F.L., D.S., D.C., and I.B. performed and analyzed penetration-depth measurements. J.F.L., S.V.T., and I.B. developed and characterized the experimental setup. N.K., R.S., and Y.O. grew samples. J.F.L., M.S., and I.B. interpreted results. I.B. devised and supervised the work in all respect. All authors discussed the results and commented the manuscript that was written by J.F.L., M.S., and I.B.

\subsection{Competing Interests} The authors declare that they have no competing financial interests.

\subsection{Correspondance} Correspondence and requests for materials should be addressed to I.B. (email: bonalde@ivic.gob.ve).


\end{document}